\documentclass[aps,prb,twocolumn,superscriptaddress,showpacs]{revtex4}
\usepackage{graphicx,bm,color,float,textcomp,mathtools}

\newcommand\sho{$\rm SrHo_2O_4$}
\newcommand\sgo{SrGd$_{\textrm 2}$O$_{\textrm 4}$}
\newcommand\seo{$\rm SrEr_2O_4$}
\newcommand\sdo{$\rm SrDy_2O_4$}
\newcommand\syo{$\rm SrYb_2O_4$}
\newcommand\slo{Sr$Ln_2$O$_4$}
\newcommand\afm{antiferromagnet}

\begin{document}
\title{Magnetic properties of geometrically frustrated \sgo}
\date{\today}

\author{O. Young}
\affiliation{Department of Physics, University of Warwick, Coventry CV4 7AL, United Kingdom}
\author{G. Balakrishnan}
\affiliation{Department of Physics, University of Warwick, Coventry CV4 7AL, United Kingdom}
\author{M. R. Lees}
\affiliation{Department of Physics, University of Warwick, Coventry CV4 7AL, United Kingdom}
\author{O. A. Petrenko}
\affiliation{Department of Physics, University of Warwick, Coventry CV4 7AL, United Kingdom}

\begin{abstract} 
A study of the magnetic properties of the frustrated rare earth oxide \sgo\ has been completed using bulk property measurements of magnetization, susceptibility and specific heat on single crystal samples.
Two zero-field phase transitions have been identified at 2.73 and 0.48~K.
For the field, $H$, applied along the $a$ and $b$ axes, a single boundary is identified that delineates the transition from a low field, low temperature magnetically ordered regime to a high field, high temperature paramagnetic phase.
Several field-induced transitions, however, have been observed with $H \parallel c$.
The measurements have been used to map out the magnetic phase diagram of \sgo, suggesting that it is a complex system with several competing magnetic interactions.
The low-temperature magnetic behavior of \sgo\ is very different compared to the other \slo\ (\emph{Ln} = Lanthanide) compounds, even though all of the \slo\ compounds are isostructural, with the magnetic ions forming a low-dimensional lattice of zigzag chains that run along the $c$ axis.
The differences are likely to be due to the fact that in the ground state Gd$^{3+}$ has zero orbital angular momentum and therefore the spin-orbit interactions, which are crucial for other \slo\ compounds, can largely be neglected.
Instead, given the relatively short Gd$^{3+}$-Gd$^{3+}$ distances in \sgo, dipolar interactions must be taken into account for this \afm\ alongside the Heisenberg exchange terms.
\end{abstract}


\maketitle

\section{Introduction}
Geometrical frustration arises in systems where the interactions between the magnetic moments are incompatible with their spatial arrangement in a lattice, so that at low temperatures not all of the interaction energies can be simultaneously minimized.~\cite{Ramirez_1994, Buschow_2001, Diep_2005}
Thus, in the presence of \afm ic exchange interactions, many magnets based on corner- or edge-sharing triangles~\cite{Ramirez_1990, Collins_1997, honey-struc} or tetrahedra~\cite{Schiffer_1994, spinels, Gardner_2010} can exhibit geometric frustration.
A general consequence of frustration is the establishment of magnetic order at temperatures much lower than what would be expected from the strength of the exchange interactions, and the absence of magnetic order down to the lowest measured temperatures.~\cite{Villain_1979, Binder_1986, Moessner_1998, Canals_1998}
While some frustrated systems remain in a highly degenerate manifold of ground states, others form a unique ground state at very low temperatures.
In this regime small perturbations to the Hamiltonian, such as further-neighbour exchange, single-ion anisotropy, magnetic dipolar interactions, etc., as well as other effects such as quantum fluctuations, become important and can lead to unusual spin arrangements.
New geometrically frustrated systems are constantly being discovered, and their behavior is often complex, with a rich variety of low-temperature properties.

\begin{figure}[tb]
   \centering
	\includegraphics[width=0.99\columnwidth]{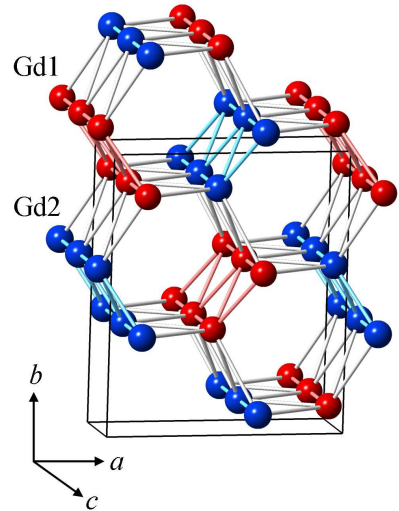}
	\caption{Positions of the magnetic ions in \sgo, with the two crystallographically inequivalent sites of the rare earth ions shown in different colors (red and blue).
	When viewed in the \emph{a-b} plane, honeycombs of the Gd$^{3+}$ ions are visible.
	Zigzag chains running along the $c$ axis connect the honeycomb layers and give rise to geometric frustration.
	The box indicates the dimensions of the crystallographic unit cell.}
   \label{Lnsublattice}
\end{figure}

The \slo\ (\emph{Ln} = Lanthanide) family of compounds,~\cite{Lopato_1973} which crystallise in the form of calcium ferrite,~\cite{Decker_1957} space group \emph{Pnam}, have been suggested as lattices that could give rise to frustrated magnetism.
For the case of \sgo, the magnetic Gd$^{3+}$ ions are linked in a network of triangles and hexagons, as shown in Fig.~\ref{Lnsublattice}, and there are two crystallographically inequivalent sites for the rare earth ions (which are shown in red and blue).
Along the $c$ axis, the magnetic ions form zigzag chains and the nearest distance between the Gd$^{3+}$ ions is about 3.48~\AA.~\cite{SGO_crystal_structure}
The zigzag chains (shown with either red or blue bonds) which consist of the Gd$^{3+}$ ions sitting in the same crystallographic positions have shorter Gd$^{3+}$-Gd$^{3+}$ separations across the zigzags (3.54 to 3.62~\AA ) as compared to the longer bonds (shown in gray, 3.85 to 4.12~\AA ), which connect the Gd$^{3+}$ ions at inequivalent positions.~\cite{SGO_crystal_structure}
The zigzag structures can be frustrated if the dominant exchange is antiferromagnetic, and are magnetically equivalent to spin chains with first- and second-nearest-neighbor interactions. 
The chains of Gd$^{3+}$ ions interconnect by forming a distorted \emph{honeycomb} structure, a bipartite lattice made up of edge sharing hexagons, in the \emph{a-b} plane.

The magnetic characterization of powder samples of the \slo\ family of compounds began with the work of Karunadasa~\textit{et al}.~\cite{Karunadasa_2005}
In this early study, measurements of the magnetic susceptibility for all of the \slo\ compounds had revealed a disparity in the measured Weiss temperatures, $\theta_{CW}$, and the lack of long-range order down to 1.8~K.\cite{Karunadasa_2005}
Single crystals of these oxides have been grown,\cite{Balakrishnan_2009,Quintero_2012} and four members of the \slo\ family, where $Ln = $Dy,\cite{Cheffings_2013} Ho,\cite{Hayes_2012,Young_2012,Young_2013,Poole_2014,Wen_2014} Er,\cite{Petrenko_2008, Hayes_2012} and Yb,\cite{Quintero_2012} have been studied in detail.
Despite the large reported values of the  Weiss temperatures ($-13.5$ to $-99.4$~K),\cite{Karunadasa_2005} measurements of the low-temperature susceptibility and heat capacity on the \sho,\cite{Hayes_2012}  \seo,\cite{Petrenko_2008} and \syo~\cite{Quintero_2012} materials revealed that their ordering temperatures are all below 1~K, and for \sdo~\cite{Cheffings_2013} no transitions to long-range order have been observed. 
All the compounds also show signs of short-range correlations in the bulk properties.
This short-range order has been further investigated using neutron diffraction measurements, and these have revealed a rich variety of low-temperature, low-dimensional magnetic behavior.\cite{Petrenko_2007, Petrenko_2008, Hayes_2011, Young_2012, Quintero_2012}
The application of a magnetic field induces a variety of transitions in all of the \slo\ compounds.\cite{Hayes_2011}
The crystals are highly anisotropic, and for \seo, and \sho\ and \sdo\ plateaux in the magnetization curves at approximately one third of the saturation value, appear for certain values of the applied field.
Such features are usually indicative of the stabilization of a colinear two-spins-up one-spin-down (uud) structure.

In this paper, we report on the low-temperature properties of \sgo, single crystals of which have been grown for the first time using the floating zone technique and examined using susceptibility, $\chi(T)$, magnetization, $M(H)$, and specific heat, $C(T)$ and $C(H)$, measurements.
The magnetic Gd$^{3+}$ ions have rather different electronic properties compared to the other lanthanides (as they are almost isotropic in the ground state where the orbital angular momentum is zero), and hence the magnetic behavior of \sgo\ is expected to be markedly different to the other members of the \slo\ series.
Our preliminary measurements of $\chi(T)$ and $M(H)$ on a powder sample of \sgo, indicated that there \textit{are} transitions in this compound at 2.73~K in $\chi(T)$ and around 19~kOe in $M(H)$, contrary to what has been reported in~Ref.~\citenum{Karunadasa_2005}.
These transitions were probably missed in the previous investigation as a result of having too large a step size in field, $H$, and temperature, $T$, during the data collection.
Low-temperature measurements of $\chi(T)$ on single crystal samples for the fields applied along each of the principal axes, show that below 2.73~K the most dramatic changes in the susceptibility occur when the field is applied along the $c$ axis.
Low-temperature $M(H)$ data indicate that the magnetization processes along the $a$ and $b$ axes are quite uneventful, but that there are several in-field transitions when $H$ is applied along the $c$ axis.
For $H \parallel a$ and  $H \parallel b$ a single boundary is identified that delineates the transition from a low field, low temperature magnetically ordered regime to a high field, high temperature paramagnetic phase.
Specific heat measurements on the single crystals of \sgo\ were first performed in zero field, and these indicate that in addition to the transition seen at 2.73~K, another transition at an even lower temperature of 0.48~K is observed.
$C(T)$ in several applied fields and $C(H)$ at a range of temperatures were also measured, and the results of all the bulk property measurements have been used to construct an \textit{H~-~T} phase diagram of \sgo.
Multiple magnetic phases have been identified, and this suggests that the magnetic ordering scheme in \sgo\ is quite complex and rather different to that of the other \slo\ compounds.
This is likely to be due to the fact that in the ground state the Gd$^{3+}$ ions have zero orbital angular momentum.
Hence, crystal-field splittings are expected to be less important in \sgo, making it the best candidate in the \slo\ series for a realisation of a classical Heisenberg exchange \afm.
Dipole-dipole interactions are expected to be the leading perturbations in the Hamiltonian.
\sgo\ thus provides an interesting comparison and furthers the study of the influence of the spin-spin and spin-orbit interaction on the physics of the \slo~\cite{Petrenko_2014} and similar systems, such as Ba$Ln_2$O$_4$.~\cite{Doi_2006, Besara_2014, Aczel_2014}

\section{Experimental Details}
\begin{table}[tb]
\begin{center}
\begin{tabular}{| p{1.5cm} | c | c |}
\hline \hline
 Atom & $x$ & $y$ \\
\hline 
Sr	& 0.7506(5) & 0.6489(4) \\
Gd1	& 0.4270(4) & 0.1127(3) \\ 
Gd2	& 0.4161(4) & 0.6110(3) \\
O1	& 0.220(3) &  0.181(2) \\
O2	&  0.135(3) &  0.479(3) \\
O3	&  0.510(3) &  0.785(2) \\
O4	&  0.423(4) &  0.420(2) \\
\hline \hline
\end{tabular}
\end{center}
\caption{Refined crystal structure parameters for \sgo\ at room temperature.
All the atoms occupy the $4c~(x, y, 0.25)$ site of the space group \textit{Pnam} with the lattice parameters $a~=~10.1321(1)$, $b~=~12.0614(1)$ and $c~=~3.47566(2)$~\AA.}
\label{sgo-riet-struc}
\end{table}

Single crystal samples of \sgo\ were grown using the floating zone method, similar to the procedure reported in~Ref.~\citenum{Balakrishnan_2009}.
Initially, powder \sgo\ samples were prepared from high purity (99.99\%) starting compounds SrCO$_3$ and Gd$_2$O$_3$, in an off-stoichiometric ratio of 1~:~0.875 (as it was found from x-ray diffraction measurements that when powders were mixed in a stoichiometric ratio the final materials contained a large, $\sim 15$\%, impurity phase of Gd$_2$O$_3$).
Following the procedure described for the earlier syntheses of \sgo,\cite{Lopato_1973, Karunadasa_2005} the powders were ground together and fired at ambient pressure in air at 1350$^\circ$C for a total of 48~hours in an alumina crucible.
In general, one intermediate grinding was used to ensure homogeneity of the mixtures.
The purity of the \sgo\ powder was verified by performing a Rietveld refinement\cite{Topas} using x-ray diffraction data collected at room temperature, and allowing for two phases in the material - the desired \sgo\ and an impurity phase of the starting compound Gd$_2$O$_3$. 
The results suggest that the \sgo\ sample that was prepared starting with an off-stoichiometric SrCO$_3$/Gd$_2$O$_3$ ratio is $99.80\%$ pure \sgo, with $\chi^2 = 1.158$ obtained for the fit.
The atomic positions for \sgo\ (within the orthorhombic space group \textit{Pnam}) are given in Table~\ref{sgo-riet-struc} together with the refined unit cell parameters.

The \sgo\ powder was subsequently isostatically pressed into rods of $\sim$7~mm diameter and $\sim$80~mm in length, and sintered in air at 1100$^\circ$C for 24 hours.
Since no previous \sgo\ crystals existed, to start the growth, a polycrystalline rod was used as a ``seed''. 
A high temperature optical furnace (Crystal Systems Inc. Optical Floating Zone Furnace Model FZT-12000-X-VI-VP) equipped with four Xe~arc lamps focused by four ellipsoidal mirrors was used.
The feed and seed rods were counter-rotated at 10-20~rpm, and growth speeds ranging from 3~to~6~mm~h$^{-1}$ were used. 
Several atmospheres were tried for the growth of \sgo, but the best results were obtained for growths carried out in air at ambient pressure.
The \sgo\ crystals grown in this manner were transparent to light, but were not wholly crack free, although large single crystal regions could be isolated.

Once the crystal was produced, several samples were aligned using a backscattering x-ray Laue method and cut into thin rectangular prisms with faces perpendicular to the principal crystal axes. 
The samples varied from 10 to 15~mg for magnetization and magnetic susceptibility to 0.14 to 12.5~mg for specific heat measurements.
Demagnetization corrections were applied following Aharoni.~\cite{demag2}

A Quantum Design SQUID magnetometer was used to measure the magnetic susceptibility and magnetization along each of the principal crystallographic directions (with the applied field within an estimated $3^{\circ}$ accuracy of the principal axes) in the ranges of $0.5 < T < 400$~K and $0 < H < 70$~kOe.
The temperature range was extended below 1.8~K using an iQuantum [IQ2000- AGHS-2RSO $^3$He system] refrigerator insert.~\cite{iQuantum}
Additional higher-field (up to 100~kOe) magnetization measurements were performed using an Oxford Instruments vibrating sample magnetometer (VSM) with the lowest achievable temperature of 1.4~K.

Specific heat measurements were performed using a Quantum Design calorimeter, with the addition of a $^3$He insert, and were carried out both in zero field in the temperature range of $0.4 < T < 400~\rm K$, and for the fields of up to 90~kOe applied along the $c$ axis.
The heat capacity of single crystal samples of SrY$_2$O$_4$ and SrLu$_2$O$_4$, which are non-magnetic and isostructural to \sgo, allowed for an estimation of the lattice contribution to the specific heat.

\section{Results and Discussion}
\subsection{Temperature dependence of the magnetic susceptibility}
\subsubsection{High-temperature limit}
\begin{figure}[tb]
   \centering
	\includegraphics[width=0.99\columnwidth]{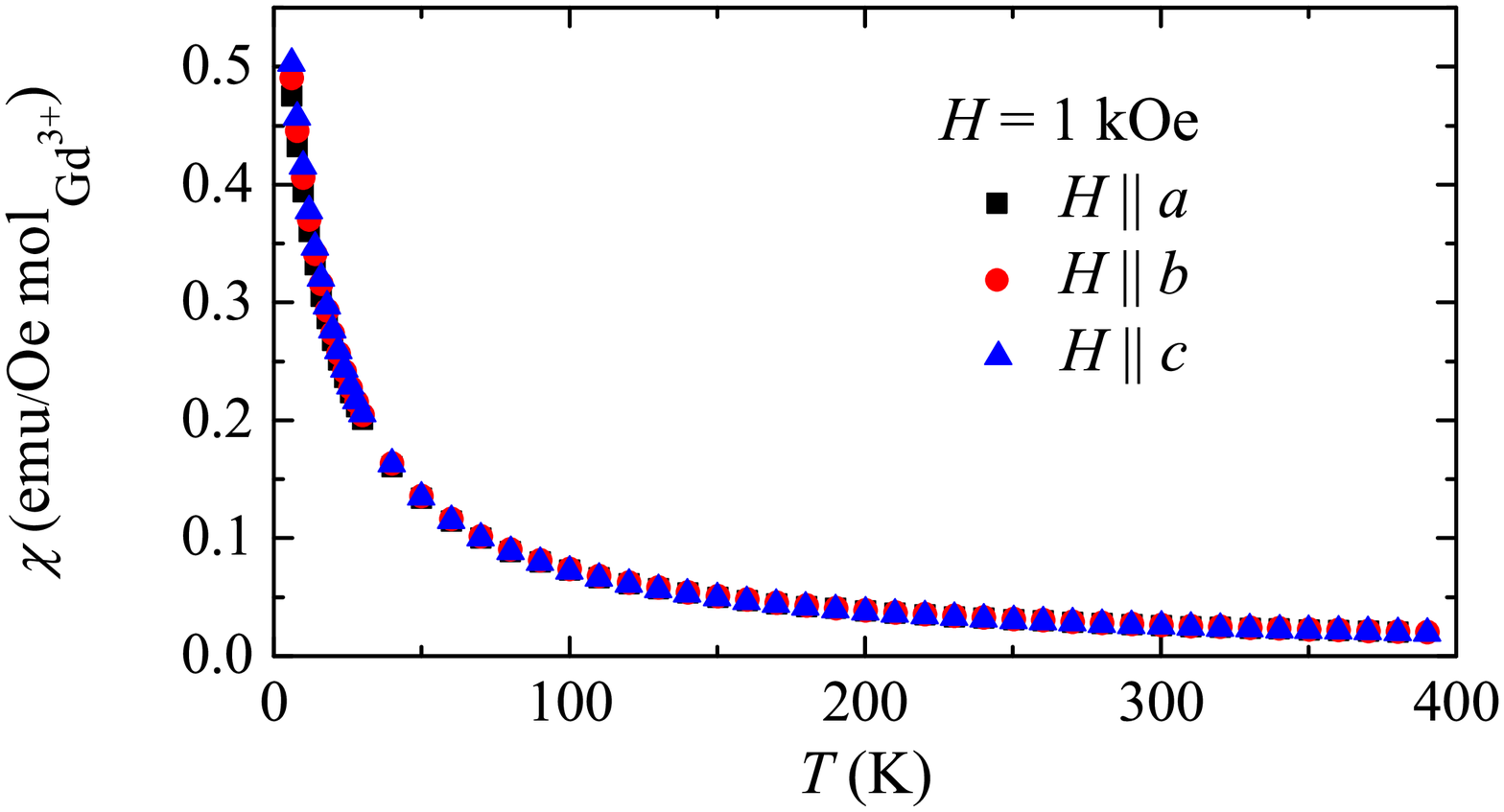}
	\includegraphics[width=0.99\columnwidth]{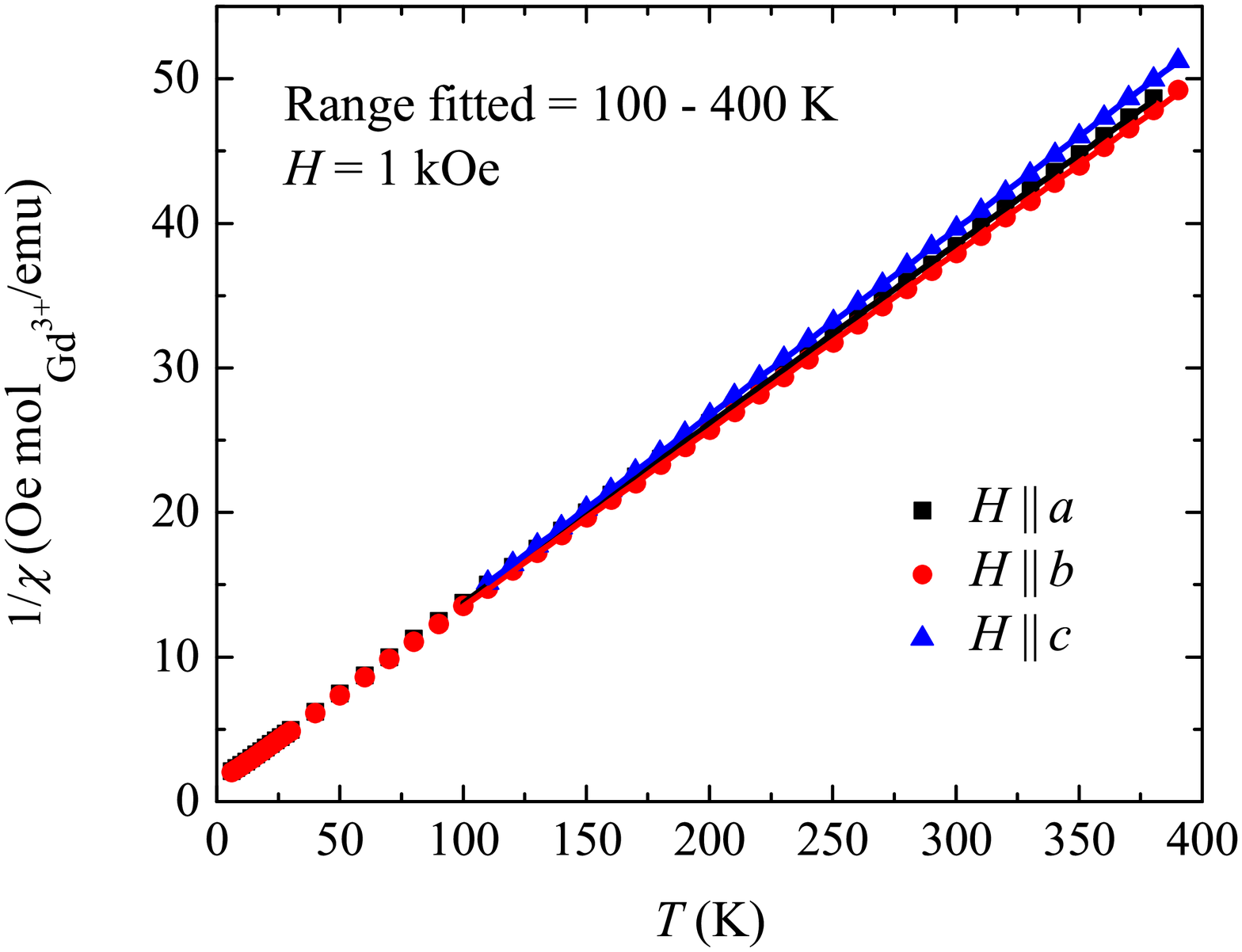}
	\caption{Top: Magnetic susceptibility versus temperature in an applied field of 1~kOe in the temperature range of 6 to 400~K for the field applied along the three principal axes of a single crystal sample of \sgo.
	Bottom: Reciprocal of the molar susceptibility versus temperature and the least-squares regression fits to the data (using the Curie-Weiss model).}
   \label{sgo-sc-HighTchi}
\end{figure}

The top panel of Fig.~\ref{sgo-sc-HighTchi} shows the magnetic susceptibility versus temperature in 1~kOe, for a field applied along each of the three principal axes for a single crystal sample of \sgo.
The bottom panel of Fig.~\ref{sgo-sc-HighTchi} presents the temperature dependence of the inverse susceptibility.
Here, the data are fitted to straight lines in the temperature range of 100 to 400~K.
The parameters of these fits, the Weiss temperatures, $\theta_{\rm CW}$, and the calculated effective moments per magnetic ion, $\mu_{\rm eff}$, are listed in Table~\ref{SGOcurieweiss}. 
The average value of $\mu_{\rm eff}$ is relatively close to the theoretical limit of 7.94~$\mu_{\rm B}$, predicted for the free ion using Hund's rules.
The average value of $\theta_{\rm CW}$ is consistent with $\theta_{\rm CW}$~=~$-10.4(1)$~K, which is obtained from our measurements made on a powder sample of \sgo, and this is comparable to the value of $-9.0(6)$~K obtained by Karunadasa,~\textit{et al}.~\cite{Karunadasa_2005}

\begin{table}[tb]
\begin{center}
\begin{tabular}{| c | c | c | c | c | c |}
\hline \hline
 & $H \parallel a$ & $H \parallel b$ & $H \parallel c$ & Mean & Powder \\
\hline 
$\mu_{\rm eff} (\mu_{\rm B})$ & 8.03(8) & 8.10(5) & 7.88(2) & 8.00(3) & 8.03(1) \\
$\theta_{\rm CW}$ (K) & $-11.0(6)$ & $-7.4(2)$ & $-12.3(1)$ & $-10.3(2)$ & $-10.4(1)$ \\
\hline \hline
\end{tabular}
\end{center}
\caption{$\mu_{\rm eff}$ and $\theta_{\rm CW}$ for the fits of the data collected with the field applied along each of the principal axes of a single crystal sample of \sgo, their average values and a comparison to the data collected for a powder sample of \sgo.
The value of the effective moment is close to 7.94~$\mu_{\rm B}$,~\cite{Kittel_1996} which is calculated using Hund's rules for a Gd$^{3+}$ ion.}
\label{SGOcurieweiss}
\end{table}
There are \emph{no} large differences in the Weiss temperatures or the effective moments for the three principal crystal directions of \sgo, contrary to what is observed for the other \slo\ compounds.\cite{Hayes_2012, Quintero_2012}
It is likely that for all the \slo\ compounds other than \sgo, the differences in the high-temperature susceptibility curves are due to the effects of low-lying crystal field levels.\cite{Poole_2014}
For \sgo, however, in the ground state Gadolinium has a pure spin magnetic moment, and thus there is no distortion of the spherical 4$f$ charge density due to the spin-orbit coupling, and hence no corresponding crystal field anisotropy.

\subsubsection{Low-temperature limit}
\begin{figure}[tb]
   \centering
	\includegraphics[width=0.99\columnwidth]{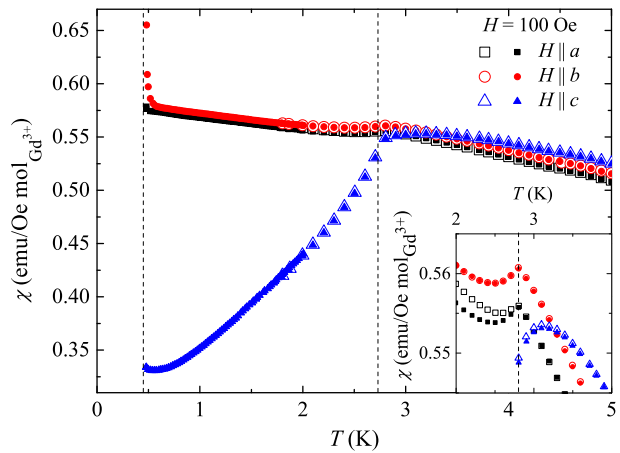}
	\caption{Magnetic susceptibility obtained in the temperature range of 0.5 to 5~K for a field 100~Oe applied along each of the three principal axes of \sgo.
	Two dashed lines represent 0.48~K and 2.73~K, where $\lambda$-anomalies were observed in heat capacity data in zero field.
	A cusp is seen in the susceptibility at 2.73~K when the field is applied along either the $a$ or $b$ axes, while a sharp decrease occurs in $\chi(T)$ for a field applied along the $c$ axis, which is highlighted in the inset.
	Also, below 2.73~K, small differences between ZFC (closed symbols) and FC (open symbols) measurements become apparent for $H \parallel a$.}
   \label{sgo-sc-LowTchi}
\end{figure}
\begin{figure}[tb]
   \centering
	\includegraphics[width=0.99\columnwidth]{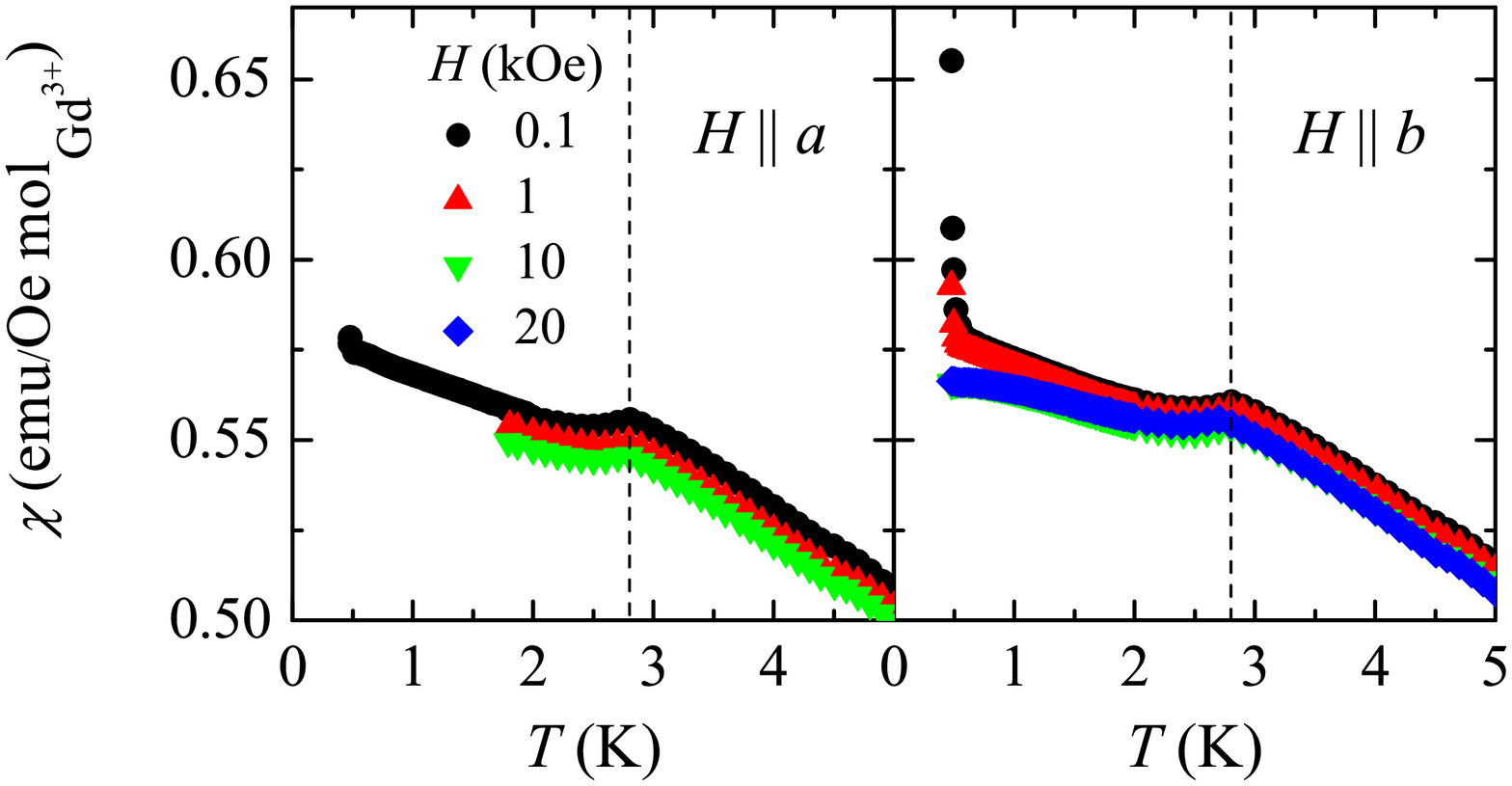}
	\includegraphics[width=0.99\columnwidth]{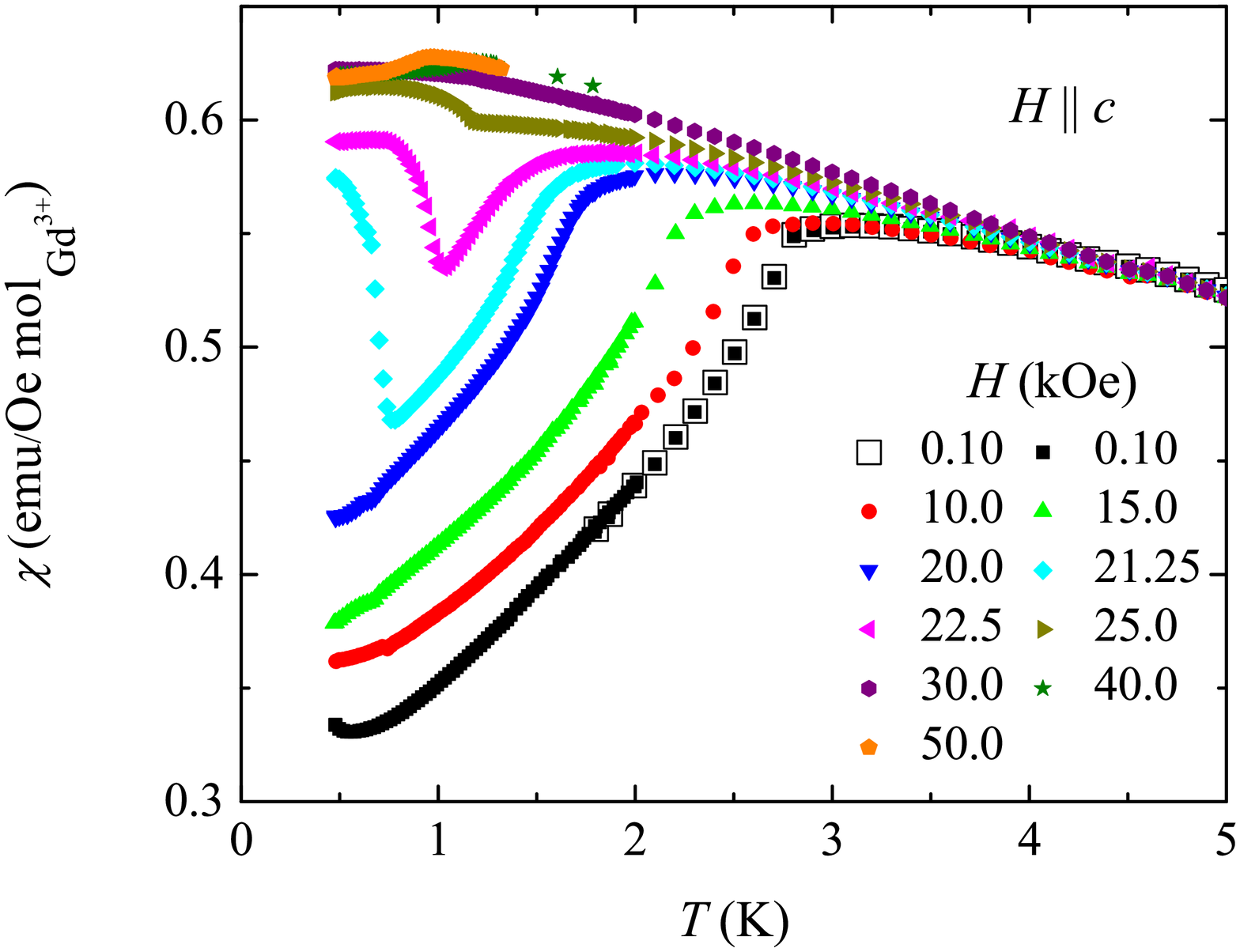}
	\caption{Top: Temperature dependence of the magnetic susceptibility in different fields applied along the (left) $a$ and (right) $b$ axes of \sgo.
	There seems to be very little difference in the behavior at low fields and upon increasing the applied field along these axes.
	Bottom: low-temperature magnetic susceptibility obtained in the temperature range of 0.5 to 5~K, and field range of 0.10 to 50~kOe, with the fields applied along the $c$ axis of \sgo.
	The transition seen at 2.73~K is suppressed by the application of higher fields, and in fields between 20 and 30~kOe a second feature in the magnetic susceptibility is observed.
	In fields above 30~kOe, again, only a single cusp in the low-temperature susceptibility is visible.}
   \label{sgo-sc-LowTchi-c}
\end{figure}

The low-temperature susceptibility measurements on the single crystal samples of \sgo\ are presented in Fig.~\ref{sgo-sc-LowTchi} for a field of 100~Oe applied along the $a$, $b$, and $c$ directions.
Here, a small cusp in the susceptibility at 2.73~K for the $a$ and $b$ axes, and a large decrease $\chi(T)$ when the field is applied along the $c$ axis suggests the presence of a phase transition, with the region of interest highlighted in the inset to Fig.~\ref{sgo-sc-LowTchi}.
This is the first indication of a transition to a long-range ordered state identified for this compound, and the phase transition temperature appears to be much higher than that recorded for the other members of the \slo\ series of compounds.
Around the transition temperature, in low applied fields such as 100~Oe, there is only a slight difference between the data obtained on warming after cooling in field (FC) and the zero-field-cooled warming (ZFC) regimes, which suggests that the magnetic susceptibility of \sgo\ is not particularly sensitive to sample history.

The lowest reachable experimental temperature when measuring magnetic susceptibility with the aid of a $^3$He probe is $\sim$0.5~K.
Unfortunately this is not low enough to observe the second transition seen at 0.48~K in zero field for a powder sample of \sgo\ using heat capacity (see Section~\ref{sec-sgo-hc-0T}).
However, the susceptibility data, acquired at low fields (such as 0.1~kOe in Fig.~\ref{sgo-sc-LowTchi}), along all of the principal axes of \sgo\ show an upturn in $\chi(T)$ at the lowest temperatures, hinting to the presence of a second transition in \sgo\ at temperatures just below 0.5~K.

The temperature dependences of the magnetic susceptibilities in a range of fields applied along the $a$ and $b$ axes of \sgo\ are shown in the top panels of Fig.~\ref{sgo-sc-LowTchi-c}.
No extra features are observed in larger fields.
The low-temperature magnetic susceptibility curves obtained in the temperature range of 0.5 to 5~K, and field range of 0.1 to 50~kOe, with the fields applied along the $c$ axis of \sgo\ are shown in the bottom panel of Fig.~\ref{sgo-sc-LowTchi-c}. 
Upon the application of field, the transition seen at 2.73~K is suppressed to lower temperatures.
In fields between 20 and 30~kOe a second feature in the magnetic susceptibility is observed, and in fields above 30~kOe, only a single cusp in the low-temperature susceptibility is visible.
This suggests a rich \textit{H~-~T} phase diagram for the fields applied along this axis of \sgo, and this will be discussed in Section~\ref{sec-sgo-sc-phaseD}.

\subsection{Field dependence of the magnetization}
\begin{figure}[tb]
   \centering
	\includegraphics[width=0.99\columnwidth]{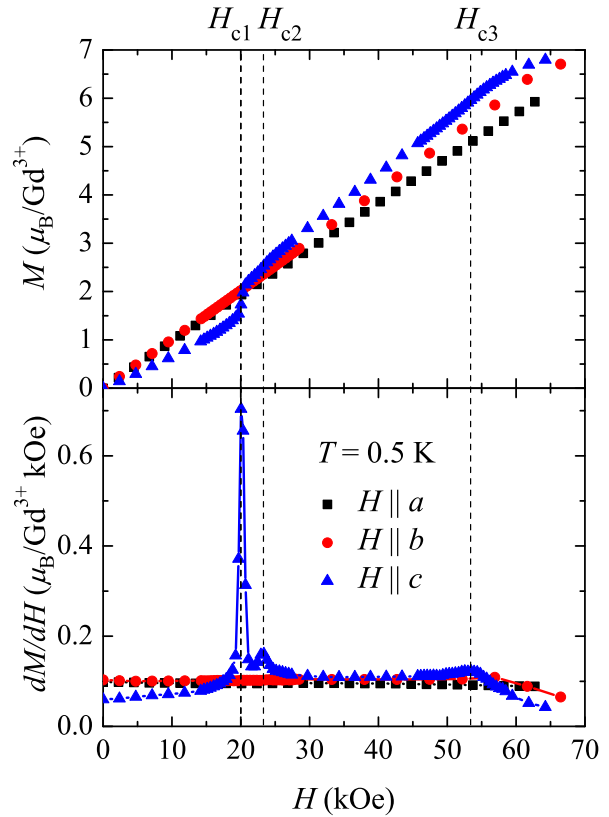}
	\caption{Top: Magnetization curves obtained with the field applied along the principal axes of \sgo\ at 0.5~K in the range of 0 to 70~kOe.
	Bottom: Field derivatives of the magnetization.
	The field induced transitions are indicated by dashed lines at $H_{\rm c1}=20.0$~kOe, $H_{\rm c2}=23.3$~kOe and $H_{\rm c3}=53.4$~kOe.
	Between $H_{\rm c1}$ and $H_{\rm c2}$, when $H \parallel c$, a narrow plateau in $dM/dH$ is observed at approximately one third of the value for the maximum moment along the $c$ axis.}
   \label{sgo-MvHabc}
\end{figure}
\begin{figure}[tb]
   \centering
	\includegraphics[width=0.99\columnwidth]{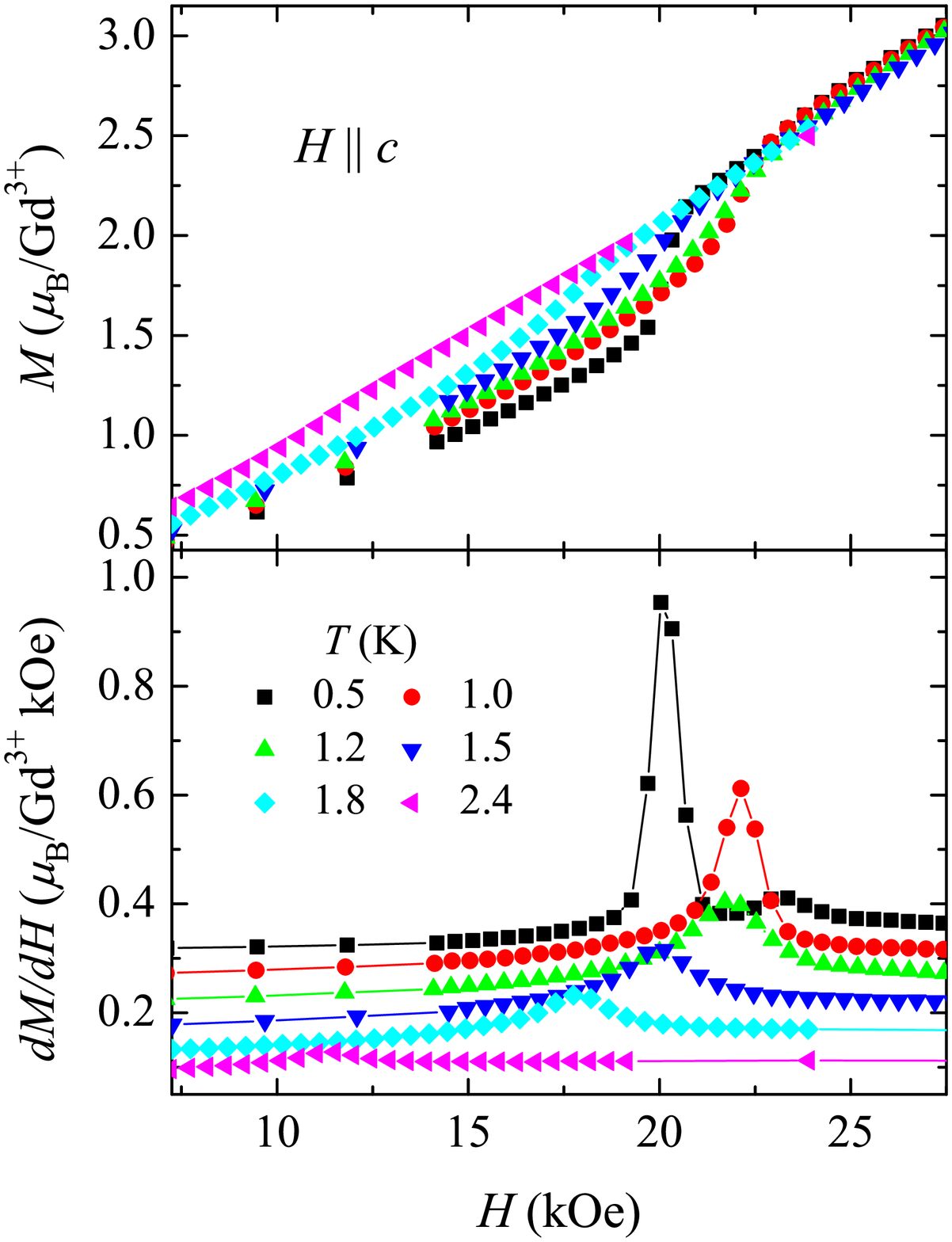}
	\caption{Top: Magnetization curves for \sgo\ obtained at several temperatures in the field range of 7.5 to 27.5~kOe, for $H \parallel c$.
	Bottom: Field derivatives of the magnetization, the curves have been offset by 0.05 units along the vertical axis.
	The double peak in the derivative of magnetization seen in the data collected at 0.5~K becomes only a single peak at higher temperatures, and this single maximum in $dM/dH$ is initially shifted up, and then down in field as the temperature is increased.}
   \label{sgo-MvHc}
\end{figure}

The field dependence of the magnetization and its derivative obtained with the field applied along the principal axes of \sgo\ at 0.5~K are shown in Fig.~\ref{sgo-MvHabc}.
In lower fields, the data collected for $H \parallel a$ shows no features in $dM/dH$ which remains small and practically flat.
A similar behavior is observed for $H \parallel b$.
In higher fields, both for $H \parallel a$ and $H \parallel b$, the magnetization deviates from a nearly straight line and shows signs of tending towards saturation, but being limited to a maximum field of 70~kOe (which after taking into account demagnetizing effects becomes even lower) it is difficult to accurately define the critical fields, therefore the error bars for the critical fields are relatively large.
However, we have been able to clearly observe the magnetic saturation transitions for the field applied along either $a$ or $b$ axes of \sgo\ at temperatures above 1.4~K taking advantage of the higher fields available in the VSM (data not shown).
Upon raising the temperature, the saturation transition moves gradually to lower fields and disappears completely above 2.73~K.
The $M(H)$ and $M(T)$ data collected using the VSM for fields applied along either the $a$ or $b$ axes are combined with the SQUID data to construct the phase diagram discussed in section \ref{sec-sgo-sc-phaseD}.

The magnetization process of \sgo\ with $H \parallel c$ (the easy axis for this material) is more complicated, with three in-field transitions seen for data collected at 0.5~K.
The initial rise in the magnetization is accompanied by a maximum in $dM/dH$ at $H_{\rm c1} \approx 20.0$~kOe, and then for a small region of the applied field the magnetization shows much slower growth (with a minimum in $dM/dH$ seen at $\sim$22~kOe), and then another small rise up to a second maximum in $dM/dH$ at $H_{\rm c2} \approx 23.3$~kOe.
Thus, it is possible that for \sgo\ $H_{\rm c1}$ and $H_{\rm c2}$ confine a narrow plateau with an average magnetization value of 2.3~$\mu_{\rm B}$ (which is equal to roughly a third of the saturation magnetization value observed with $H \parallel c$).
Such a plateaux can be a sign of a field induced stabilization of a colinear \textit{two-spins-up-one-spin-down} (uud) magnetic structure, in which on each triangle of spins, two are pointing up along the field and the third spin pointing down anti-parallel to the field direction.
Similar, albeit more distinct, plateaux have been observed in the magnetization curves of the other \slo\ compounds,~\cite{Hayes_2012} and thus, the stabilization of the uud spin structure could be a common feature of this family of materials.
A third field induced transition for $H \parallel c$ is observed at $H_{\rm c3} \approx 53.4$~kOe, and in even higher applied fields the magnetization shows signs of approaching saturation.
The demagnetization corrections only have a small impact on the position of the field-induced phase transitions at $H_{\rm c1}$ and $H_{\rm c2}$ (of under 2$\%$), but a somewhat larger effect on the position of $H_{\rm c3}$ (of $\sim5\%$).
These field induced transitions have been confirmed across multiple independently aligned \sgo\ samples and their positions in field remain robust.
Also, no hysteresis with the applied field along any of the principal crystal axes of \sgo\ is observed, as the magnetization data collected upon increasing and decreasing field coincides. 

To further investigate the region of the applied field around $H_{\rm c1}$ and $H_{\rm c2}$ with $H \parallel c$ for \sgo, $M(H)$ data was collected at several temperatures, and this and the derivative of the magnetization are shown in Fig.~\ref{sgo-MvHc}.
The double peak in the derivative of magnetization seen in the data collected at 0.5~K becomes only a single peak at higher temperatures, and this single maximum in $dM/dH$ is initially shifted up, and then down in field as the temperature is increased.
Since the two peaks in $dM/dH$ at $H_{\rm c1}$ and $H_{\rm c2}$ split from a single peak as the temperature is lowered, and with the extra data points obtained for plotting out the phase diagram for $H \parallel c$ (see Section~\ref{sec-sgo-hc-0T}), it seems plausible that at even lower temperatures the uud state would govern a larger region of the applied field.

The magnetization processes in all of the \slo\ compounds are highly anisotropic.
In \seo, the magnetic anisotropy is of pronounced easy-plane type,\cite{Balakrishnan_2009,Hayes_2012} while in \sho\ and \sdo\ an Ising-type behavior is observed (with different directions of an easy-axis on two different crystallographical sites for the magnetic ions).\cite{Hayes_2012,Poole_2014}.
In \syo, the magnetization measured with $H\parallel c$ is much higher than in perpendicular directions, but the magnetic moments seem to have significant components along all three main symmetry axes.\cite{Quintero_2012}
In \sgo, which is not affected by crystal field splitting, almost all of the non-trivial magnetic behavior occurs for fields applied along the $c$ axis and it is possible that this direction is an effective easy-axis of magnetization.
The fact that low-temperature magnetization below $H_{\rm c1}$ is significantly large (well above 1~$\mu_{\rm B}$ per Gd$^{3+}$) implies that the low-field magnetic structure in \sgo\ is noncolinear.

An interesting parallel to the behavior of \sgo\ may be found by looking at the Gd pyrochlore titanate.
A Heisenberg-type \afm, $\rm Gd_2Ti_2O_7$ orders magnetically at about 1~K.\cite{Raju_1999}
Despite the absence of a significant single ion anisotropy for the Gd$^{3+}$ ions, $\rm Gd_2Ti_2O_7$ shows a highly anisotropic behavior in applied fields at lower temperatures,\cite{Petrenko_2004} with multiple field-induced transitions.~\cite{Petrenko_2012}
In Section~\ref{sec-sgo-sc-phaseD}, we will further extend the analogy between the high-field behavior of \sgo\ and $\rm Gd_2Ti_2O_7$, after first discussing the relevant $C(H)/T$ results below.
\subsection{Specific heat}
\label{sec-sgo-hc-0T}
\begin{figure}[tb]
   \centering
	\includegraphics[width=0.99\columnwidth]{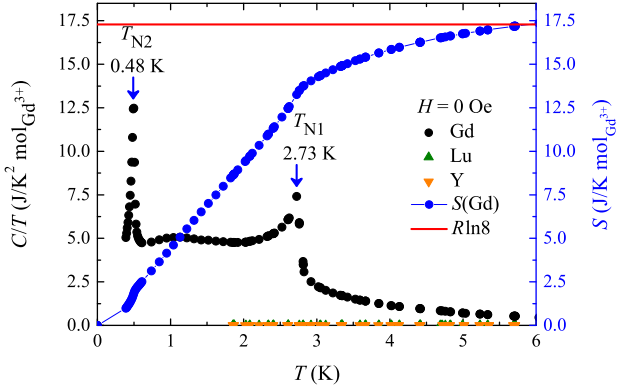}
	\caption{Temperature dependence of the specific heat divided by temperature of \sgo\ and of the nonmagnetic isostructural compounds SrLu$_2$O$_4$ and SrY$_2$O$_4$ measured in zero field.
	Two $\lambda$-anomalies are observed at $T_{\rm N1}=2.73$~K and $T_{\rm N2}=0.48$~K.
	The entropy is calculated as the area under the $C(T)/T$ curve which has been linearly extended to zero at $T$~=~0~K, with the scale given on the righthand axis.
	By 6~K, the full magnetic contribution to the entropy expected for the Gd$^{3+}$, $J=S=7/2$, system is recovered, as indicated by the solid line positioned at $R \ln{8}$.}
   \label{sgo-sc-hc-h=0}
\end{figure}

The temperature dependence of the specific heat, $C(T)$, of single crystals of \sgo\ and of the nonmagnetic isostructural compounds SrLu$_2$O$_4$ and SrY$_2$O$_4$ were measured in zero field.
The $C(T)/T$ curves are shown in Fig.~\ref{sgo-sc-hc-h=0}.
Two $\lambda$-anomalies which correspond to transitions to long-range magnetic order are observed at $T_{\rm N1}=2.73$~K and $T_{\rm N2}=0.48$~K. 
The lattice contribution to the specific heat of \sgo\ can be estimated by measuring the heat capacity of two single crystal non-magnetic isostructural compounds: SrY$_2$O$_4$ and SrLu$_2$O$_4$.
At temperatures below $\sim 6$~K, the lattice contribution to the specific heat of \sgo\ is negligible compared to the magnetic contribution, and thus the magnetic entropy can be estimated by integrating the $C(T)/T$ curve which has been extended linearly down to $T=0$~K.
It should be noted that, for a powder sample, measurements of the specific heat were also made using a dilution insert for the PPMS which extended the temperature range of the zero field measurements down to 0.05~K (data not shown).
These measurements confirmed that below $T_{\rm N2}$ $C(T)/T \rightarrow 0$ in a linear fashion as $T \rightarrow 0$~K, and that no more low-temperature transitions are present in the material down to 0.07~mK.
The entropy curve for the single crystal measurements is also shown in Fig.~\ref{sgo-sc-hc-h=0}, and the (blue) righthand axis should be used for the appropriate scale.
It appears that by 6~K the full magnetic entropy expected for the Gd$^{3+}$, $J~=~7/2$, system is recovered.

\begin{figure}[tb]
   \centering
	\includegraphics[width=0.99\columnwidth]{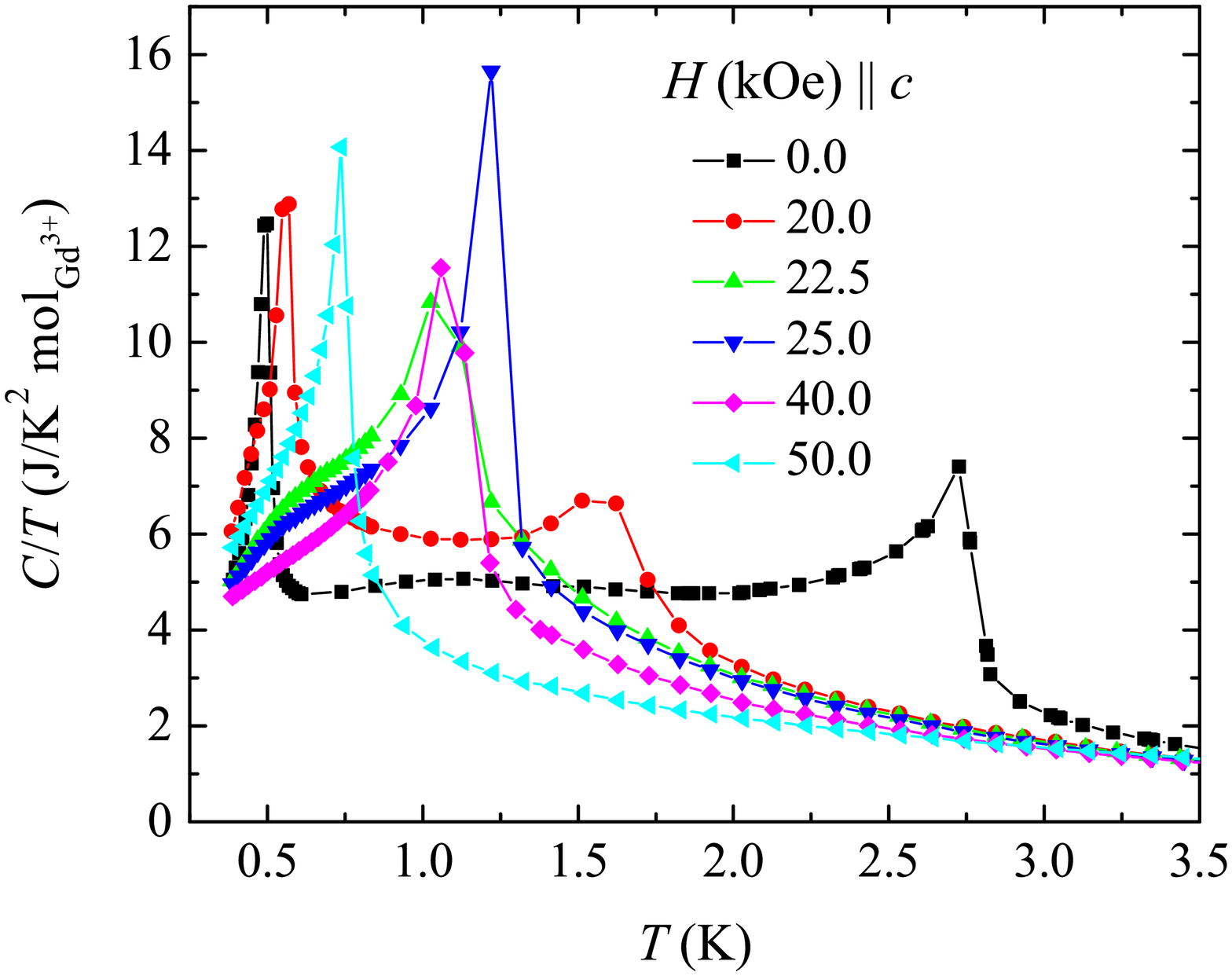}
	\includegraphics[width=0.99\columnwidth]{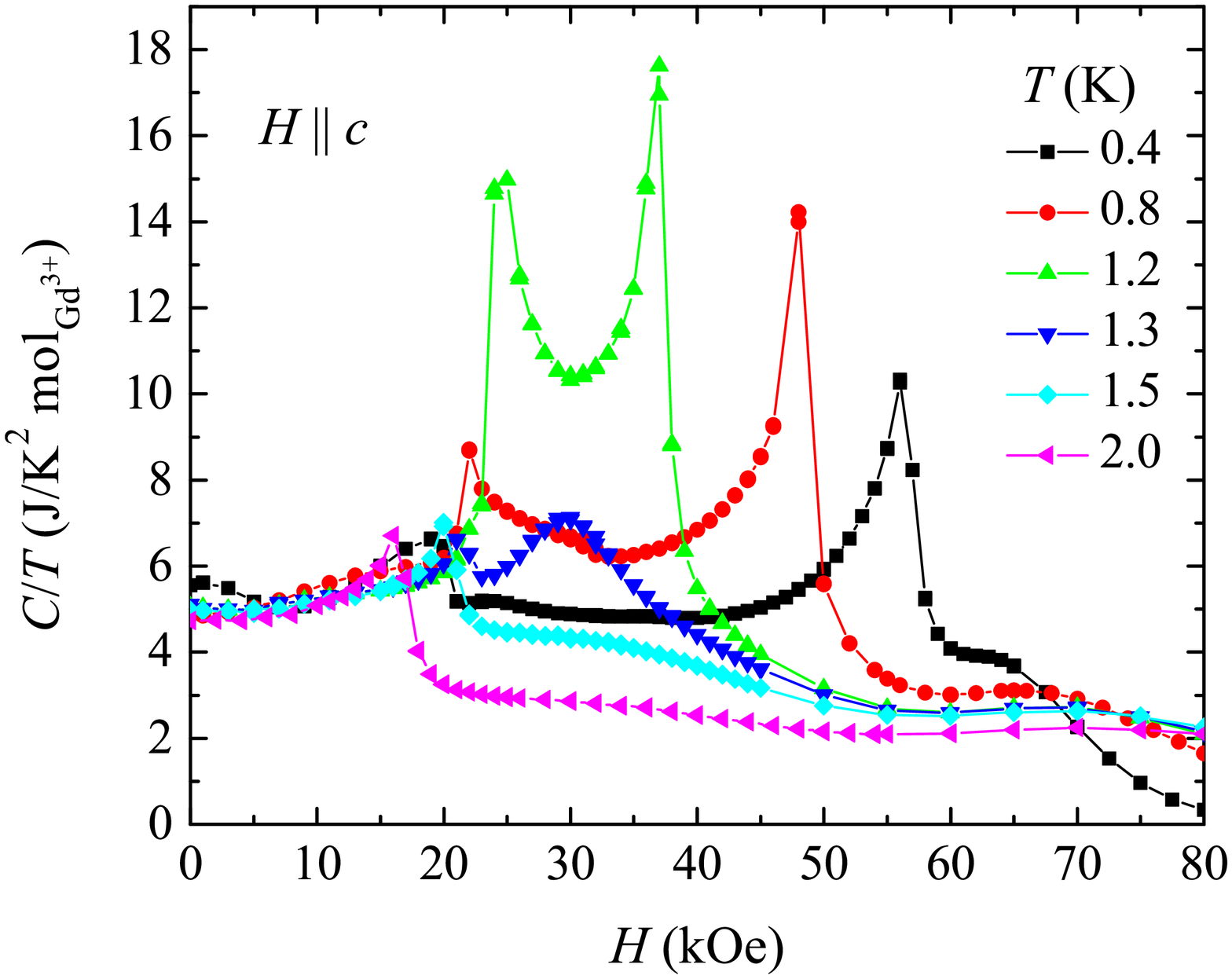}
	\caption{Top: Temperature dependence of the specific heat divided by temperature of \sgo\ in several fields for $H \parallel c$.
	The two peaks associated with the transitions move closer with higher applied fields, before merging and then this single peak is shifted to lower temperatures in the highest applied fields.
	Bottom: Field dependence of the specific heat divided by temperature at several temperatures for $H \parallel c$.
	At the lowest temperature, two peaks are observed in $C(H)$.
	As the temperature is raised, the peaks move closer together and eventually merge.
	For even higher temperatures, this single peak is suppressed and moves to lower fields.}
   \label{sgo-HC_H//C}
\end{figure}

From magnetic susceptibility and magnetization data it has become apparent that for the fields applied along the $a$ and $b$ axes, \sgo\ shows no signs of any additional field-induced transitions apart from the transition from an ordered phase into a fully polarized state.
With this in mind, the specific heat was measured for the fields applied along the $c$ axis in a broad range of both fields and temperatures.
The data for $C(T)/T$ in several applied fields are shown in the top panel of Fig.~\ref{sgo-HC_H//C}.
The two phase transitions in \sgo, whose temperature is defined by the $\lambda$ anomalies, seen in $H = 0$~Oe move closer together upon increasing the strength of the applied field.
The transitions `merge' into a single broad peak seen when $C(T)$ is measured in 22.5~kOe, and after subsequent increases of the field this single peak is first shifted to higher and then lower temperatures.
$C(H)/T$ was also measured for $H \parallel c$, and the data is shown in the bottom panel of Fig.~\ref{sgo-HC_H//C}.
At the lowest measured temperature, two peaks are visible in the heat capacity data. 
As  the temperature is increased, these two peaks move closer together, and in a small temperature range (between 1.2 and 1.3~K) are quickly suppressed in intensity before merging together at temperatures around 1.5~K.
Subsequent increases in the temperature just shift this single peak to lower fields.
All of the specific heat data collected for $H \parallel c$ are also used to construct the $H \parallel c$ phase diagram for \sgo, which is described below.

The results of heat capacity measurements as a function of both temperature and magnetic field give direct information about the boundaries of the phase transitions observed in \sgo, and these measurements demonstrate that \sgo\ is very different to other \slo\ compounds. 
For \sgo, magnetic susceptibility and specific heat measurements indicate that at least two transitions take place in zero-field at low temperatures, at 2.73 and 0.48~K, and no other \slo\ compound measured so far shows two separate transitions to different long-range orders in zero applied field.
Specific heat measurements, on \sdo,\cite{Cheffings_2013} \sho,\cite{Ghosh_2011} \seo,\cite{Petrenko_2008} and \syo \cite{Quintero_2012} indicate the presence of short-range correlations developing at temperatures much higher than the observed transition temperatures in these materials.
In the specific heat data collected for \sgo, however, short-range correlations are limited to a small temperature range just above the transition temperature of 2.73~K.
In contrast to \sgo, the magnetic entropy contribution recovered at relatively low temperatures in \seo,\cite{Petrenko_2008} \syo,\cite{Quintero_2012} and \sdo~\cite{Cheffings_2013} amounts to $R \ln{2}$, while in \sho\ it is $R \ln{5}$,\cite{Ghosh_2011} although this value is obtained over a much wider temperature range and without subtracting the nuclear Schottky anomaly.

\subsection{\textit{H~-~T} phase diagram of \sgo}
\label{sec-sgo-sc-phaseD}
\begin{figure}[tb]
   \centering
	\includegraphics[width=0.99\columnwidth]{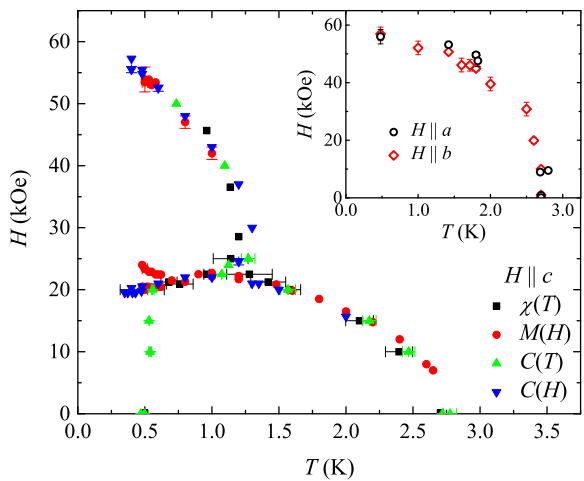}
	\caption{Main panel: Magnetic \textit{H - T} phase diagram for \sgo\ with the field applied along the $c$ axis constructed from the susceptibility, magnetization and specific heat curves.
	Inset: Field dependence of the upper critical temperature, $T_{\rm N1}$, as observed in magnetization $M(H)$ and $M(T)$ measurements for $H \parallel a$ and $H \parallel b$.
	The methods used to obtain the phase boundaries are given in Ref.~\citenum{forPD}.}
   \label{sgo-phaseD}
\end{figure}

By collating all of the susceptibility, magnetization, and specific heat data for $H \parallel c$, a magnetic field\textendash temperature phase diagram may be constructed for \sgo, and this is shown in main panel of Fig.~\ref{sgo-phaseD}.~\cite{forPD}
The phase diagram reveals the rich and complex magnetic behavior of \sgo\ when $H \parallel c$. 
In addition to the paramagnetic regime, four other phases may be identified by looking at the boundaries on the diagram.
This behavior is contrasted with a somewhat simpler response observed on application of magnetic field in directions orthogonal to $c$.
For $H \parallel a$  and $H \parallel b$, the inset to Fig.~\ref{sgo-phaseD} shows how the upper-temperature transition $T_{\rm N1}$ gradually decreases in increasing fields.
In this case, the magnetization measurements (limited to temperatures above 0.5~K) seem to indicate a single transition from a magnetically ordered to a disordered phase.
It might be interesting to extend the magnetization measurement to the temperatures below $T_{\rm N2}=0.48$~K and check how the lowest-temperature phase responds to the application of field along all there crystallographical axes.

For all three directions of an applied field, the value of the upper-field transition $H_{\rm c3}$ extrapolated to zero temperature returns approximately 60~kOe.
Treated as a saturation field, this value could potentially be used to estimate the strength of the \afm ic exchange interactions $J_{ij}$ provided that the number of interacting neighbors is known.
In the absence of such information and because of the possible competition between different exchange links involved, neither $H_{sat}$ nor the $\theta_{\rm CW}$ can be reliably converted to estimate $J_{ij}$, but what can be stated is that the energy of the exchange interactions in \sgo\ amounts to just a few kelvin.  
If this is the case, then it is rather obvious that given the relatively short Gd$^{3+}$-Gd$^{3+}$ separation (3.48~\AA\ along the $c$ axis) and a large magnetic moment ($J~=~7/2$), dipolar interactions, which only for the nearest neighbors amount to roughly a kelvin in units of temperature, cannot simply be disregarded -- they must be taken into account as a leading perturbation to a purely exchange Hamiltonian.

In zero field, two transitions take place in \sgo.
The magnetic structures of the different phases remain unknown, but at least for the intermediate temperature regime a non-zero susceptibility and magnetization rule out a simple colinear ordering. 
With the application of field along the $c$ axis of \sgo, at the lowest temperature, three field-induced transitions are observed at  $H_{\rm c1}$, $H_{\rm c2}$ and $H_{\rm c3}$.
It remains to be confirmed that the phase between $H_{\rm c1}$ and $H_{\rm c2}$ corresponds to uud order observed in other \slo\ compounds.\cite{Hayes_2012}

An interesting observation to make is on the nature of the high-field phase (above $H_{\rm c3}$) at low temperatures usually labeled as a ``fully-polarized state".
The magnetization continues to grow at a significant rate above $H_{\rm c3}$ (see Fig.~\ref{sgo-MvHabc}) and the magnetic heat capacity is also far from zero (the bottom panel in Fig.~\ref{sgo-HC_H//C} shows that the heat capacity approaches zero only in much stronger fields around 80~kOe).
This behavior is reminiscent of what has been observed in another Gd$^{3+}$ containing compound, the pyrochlore titanate.
In particular, the field dependence of the heat capacity divided by temperature observed in $\rm Gd_2Ti_2O_7$ (see Fig.~3 in Ref. \onlinecite{Petrenko_2004}) seems to be remarkably similar to what is shown in Fig.~\ref{sgo-HC_H//C} for \sgo. 
One could argue that such behavior should be attributed to the influence of dipole-dipole interactions which tilt the magnetic moments away from the field direction even in a nominally fully-polarized state.

The results of the bulk property measurements on \sgo\ would merit further experimental study using neutron diffraction.
However, such experiments are challenging to carry out due to the large absorption cross-section of naturally abundant Gd.~\cite{Sears}
To date, only a few preliminary single crystal neutron diffraction measurements have been performed on \sgo, using the D9 instrument at the Institut Laue-Langevin, Grenoble, France in zero applied field.~\cite{Young_SGO_2013}
These indicate that the magnetic order that appears below 2.73~K is commensurate with the lattice, and can be indexed with the propagation vector \textbf{k}~=~0.
Other members of the \slo\ family, such as \seo\ and \sho, also develop \textbf{k}~=~0 order (or partial order in the case of Ho) involving half of the magnetic \textit{Ln} ions which occupy the same crystallographical position, while the other half of the $Ln$ ions form a completely different magnetic arrangement.~\cite{Petrenko_2008,Hayes_2011,Young_2012,Young_2013}
Thus a \textbf{k}~=~0 \afm ic order seems to be a common feature in this family of compounds, but it remains to be seen if the lower-temperature transition found in \sgo\ at 0.48~K is related to an ordering of the second half of the Gd ions or is it a further readjustment of the \textbf{k}~=~0 phase.

\section{Conclusions}
The first high quality single crystals of \sgo\ have been grown using the floating zone technique, and these have been investigated by the specific heat, magnetization and susceptibility measurements.
\sgo\ orders magnetically at 2.73~K, a temperature lower than the measured Weiss temperatures, $\theta_{\rm CW}=-10.4(1)$~K, with a further transition taking place at 0.48~K.
The data reveal an anisotropic nature of the low-temperature magnetic structure, with the $c$ axis found to be the easy axis in the system.
At the lowest accessible experimental temperatures, three field induced transitions have been observed in \sgo\ for $H \parallel c$.
It may be conjectured that one of the phases stabilised with the applied field is an up-up-down spin order such as that seen in the other members of the \slo\ family.
For the magnetic field applied along the $c$ axis, the magnetic phase diagram of \sgo, as a function of field and temperature, was carefully mapped out.

The data illustrate that there is a large difference between the magnetic behavior of \sgo\ and that of the \sho, \seo\ and \sdo\ compounds investigated previously, even though the positions of the magnetic ions and the strength of the exchange interactions are similar.
This is not totally unexpected as the low-temperature properties of \sgo\ have to be quite different compared to the other \slo\ compounds in which the spin-orbit coupling and crystal field anisotropies play a much more important role.
An interesting point for further research is to establish the hierarchy of the magnetic interactions in \sgo, in particular the relationship between the exchange interactions and the dipolar forces and their influence on the selection of the ground state in this frustrated \afm.

\section*{ACKNOWLEDGMENTS}
We are grateful to B.~Z.~Malkin and M.~L.~Plumer for valuable discussions.
The magnetometer used in this research was obtained through the Science City Advanced Materials project: Creating and Characterising Next Generation Advanced Materials, with support from Advantage West Midlands (AWM) and was part funded by the European Regional Development Fund (ERDF). 
The authors acknowledge financial support from the EPSRC, U.K. under grant EP/I007210/1.

\bibliography{OY_SGO_refs}

\begin{thebibliography}{43}
\bibitem[{SGO()}]{SGO_crystal_structure}
\bibinfo{note}{The Gd$^{3+}$-Gd$^{3+}$ bond lengths are derived from the
  refined structural parameters presented in Section II.}
\bibitem[{for()}]{forPD}
\bibinfo{note}{In order to consistently determine the critical fields and
  temperatures, when well defined peaks were not observed in $\chi(T)$,
  $dM/dH$, $C(T)$ or $C(H)$, (for example, where the transition was observed as
  a broad peak in susceptibility), the transition point was taken to be at the
  maximum in $d(\chi T)/dT$. Also, if the transition was observed as a change
  in the gradient of the measured bulk property (an example for the
  susceptibility would be the dataset collected in 30~kOe, and for the
  magnetization measurements this was the method for determining the position
  of $H_{\rm c3}$), linear fits to the data were made on either side of the
  change in the gradient, and the transition point was defined to be where
  these two lines cross.}
\end{thebibliography}
\end{document}